\documentclass[
 aps,pra,preprint,groupedaddress,
 amsmath,amssymb,
 floatfix,
]{revtex4-2}

\usepackage{graphicx}
\usepackage{dcolumn}
\usepackage{bm}
\usepackage[linesnumbered,ruled,vlined]{algorithm2e}
\usepackage{hyperref}
\hypersetup{hidelinks}
\begin{document}

\title{Quantum-enhanced ghost imaging recognition via joint optimization of speckle patterns and quantum network parameters}

\author{Yirui Mao}
\affiliation{Institute of Signal Processing and Transmission, Nanjing University of Posts and Telecommunications, Nanjing 210003, Jiangsu, China}
\author{Xiangyu Ge}
\affiliation{Institute of Signal Processing and Transmission, Nanjing University of Posts and Telecommunications, Nanjing 210003, Jiangsu, China}
\author{Yuhang Tu}
\affiliation{Institute of Signal Processing and Transmission, Nanjing University of Posts and Telecommunications, Nanjing 210003, Jiangsu, China}
\author{Anqi Zhang}
\affiliation{Institute of Signal Processing and Transmission, Nanjing University of Posts and Telecommunications, Nanjing 210003, Jiangsu, China}
\author{Le Wang}
\affiliation{Institute of Signal Processing and Transmission, Nanjing University of Posts and Telecommunications, Nanjing 210003, Jiangsu, China}

\author{Shengmei Zhao}
\email{zhaosm@njupt.edu.cn}
\affiliation{Institute of Signal Processing and Transmission, Nanjing University of Posts and Telecommunications, Nanjing 210003, Jiangsu, China}
\affiliation{Key Lab of Broadband Wireless Communication and Sensor Network Technology, Ministry of Education, Nanjing University of Posts and Telecommunications, Nanjing 210003, Jiangsu, China}

\date{\today}

\begin{abstract}
Ghost imaging enables nonlocal image reconstruction and exhibits strong robustness against interference, but achieving high-fidelity recognition at ultra-low sampling rates remains challenging. Quantum machine learning offers a novel approach for efficient feature extraction on noisy medium-scale quantum devices; however, existing methods generally suffer from low recognition accuracy and weak noise resistance. This paper proposes a ghost imaging recognition method based on the simultaneous optimization of speckle patterns and quantum network parameters. By leveraging the mathematical equivalence between classical convolution and speckle-object dot product operations in ghost imaging, a speckle consistency regularization mechanism is introduced to achieve end-to-end joint optimization of optical coding and quantum feature extractors. A parallel 8-qubit quantum circuit employing block coding and a star-shaped entanglement structure is designed to extract higher-order features from bucket signals. Simulation results on the MNIST and Fashion-MNIST datasets show that at an ultra-low sampling rate of 1.5625\%, the proposed framework achieves recognition accuracies of 90.1\% and 81.7\%, respectively, representing a 2.6\% improvement over classical convolutional neural networks and a maximum improvement of 14.2\% over traditional hybrid quantum machine learning models. This method also exhibits strong robustness to quantum noise and has been validated on a real optical ghost imaging system, achieving an average recognition accuracy of 84.8\%. These results confirm that the joint optimization of speckle patterns and quantum network parameters provides a reliable and practical solution for low-sampling ghost imaging recognition.
\end{abstract}

\maketitle

\section{Introduction}

Ghost imaging (GI) is an imaging technique that reconstructs target information by utilizing the correlation between a reference light field and a single-pixel barrel detection signal. It is fundamentally different from traditional imaging methods that rely on direct spatial resolution \cite{strekalov1995observation,pittman1995optical,bennink2002two}. Computational ghost imaging (CGI) further simplifies the optical system architecture by using programmable spatial light modulators such as digital micromirror devices (DMDs) \cite{shapiro2008computational}. Due to its nonlocal imaging characteristics and strong robustness to environmental disturbances, GI has a wide range of applications in image recognition, lidar, optical encryption, and biological microscopy \cite{howland2011photon,yang2023underwater,yang2024ghost,darafsheh2012optical,ordonez2024single,NJYD202103009}.

In recent years, relying on the continuously improving intelligent information processing technology, the capability boundaries of GI have also been expanded. For example, attention-enhanced computational ghost imaging has significantly improved the reconstruction quality in unfamiliar spaces and underwater environments\cite{chen2025attention}; optical chaotic temporal ghost imaging can achieve high-speed information reconstruction under low-bandwidth detection conditions\cite{zhang2026optical}. With the increasingly close integration between GI and information science, more and more computational strategies are being used to overcome physical limitations.

Currently, CGI-based recognition tasks still face a long-term challenge: how to achieve high-fidelity performance under ultra-low hot sampling rates (i.e., extremely limited number of bucket probes). One existing solution is to introduce classical deep learning. For example, some studies have used convolutional neural network (CNN) to directly perform image recognition from bucket signals generated by random speckle patterns \cite{lyu2017deep}; more recent work has optimized speckle patterns and CNN parameters end-to-end, so that the illumination mode can be directly adapted to downstream tasks, thereby improving recognition performance \cite{he2025ghost}. Nevertheless, when dealing with very small amounts of measurement data, classical CNN methods still tend to saturate at low sampling rates because they are difficult to capture high-order global correlation features.

Quantum machine learning (QML) offers a promising alternative, enabling efficient feature extraction in high-dimensional Hilbert spaces using noisy medium-scale quantum (NISQ) devices\cite{pan2022quantum,lamichhane2025quantum,xu2022scalable}. Hybrid classical-quantum variational circuits have been successfully applied to pattern recognition \cite{tacchino2020quantum,henderson2020quanvolutional} and classification tasks \cite{zhang2023single,havlivcek2019supervised,hur2022quantum,li2022recent}. Recent studies have further explored the combination of QML and GI: Xiao \emph{et al.} demonstrated the practical advantages of quantum machine learning in ghost imaging, proving that quantum methods have better sample complexity than classical methods in simplified GI scenarios \cite{xiao2023practical}; Zhai \emph{et al.} proposed a quantum neural compressed sensing algorithm for ghost imaging by reparameterizing the inverse problem using variational quantum circuits \cite{zhai2025quantum}. More broadly, quantum self-attention neural networks have also been used in classification task research recently \cite{li2024quantum}, and related academic journals have published special issues focusing on quantum information processing \cite{zhang55special}, reflecting the positive role of information science in promoting the development of quantum-enabled imaging and sensing.

However, existing QML-assisted GI methods still have two major limitations: (i) the recognition accuracy drops sharply at low sampling rates due to the limited feature extraction capability of quantum networks; and (ii) the performance deteriorates rapidly under quantum noise interference.

To address the aforementioned issues, this study proposes an end-to-end ghost imaging recognition method that jointly optimizes speckle patterns and quantum network parameters. Inspired by speckle optimization strategies and quantum feature extraction methods, we unify optical coding and quantum learning within a single differentiable computational flow. By introducing a speckle consistency regularization term, we force classical convolution operations and speckle-object dot product operations to be physically equivalent, thereby achieving synchronous updates of illumination patterns and quantum circuit parameters. This joint optimization strategy not only enhances feature extraction capabilities at ultra-low thermal sampling rates but also improves robustness to quantum noise.

The main contributions of this paper are as follows. (1) We establish a speckle-quantum joint optimization framework that introduces a speckle consistency regularization for synchronous optimization of speckle patterns and quantum parameters, effectively bridging physical priors with data-driven learning. (2) A hardware-efficient parallel quantum circuit with patch-based encoding and star-topology entanglement is designed for high-order feature extraction under NISQ constraints. (3) At a sampling rate of $1.5625\%$, the proposed framework achieves $90.1\%$ and $81.7\%$ accuracy on MNIST and Fashion-MNIST, respectively, outperforming classical CNNs by 2.6 percentage points and unconstrained hybrid QML models by up to 14.2 percentage points. (4) On a real optical GI system, it attains an average accuracy of $84.8\%$, with over $90\%$ for simple-structured digits (0 and 7), confirming its practicality under realistic noise conditions.

The remainder of this paper is organized as follows. Section~\ref{sec:method} presents the proposed ghost recognition method, including the speckle--quantum joint optimization framework and the design of the quantum neural network. Section~\ref{sec:results} reports numerical simulation and optical experimental results, followed by discussions. Section~\ref{sec:conclusion} concludes the paper.

\section{Proposed Ghost Recognition Method}
\label{sec:method}

\subsection{The Recognition method}

\begin{figure}[!t]
\includegraphics[width=.7\textwidth]{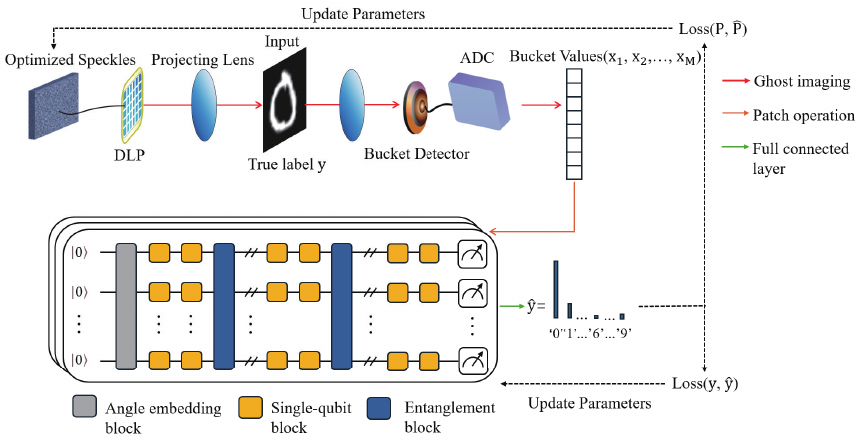}
\caption{\label{FIG1}Schematic diagram of the proposed ghost recognition method.}
\end{figure}

Fig.~\ref{FIG1} illustrates the proposed ghost recognition method. Speckles are modeled as trainable convolution kernels, and a variational quantum mapping is adopted to implicitly extract target-relevant features. Speckle structures and quantum feature spaces are optimized synchronously under a unified objective, enabling deep integration of physical priors and data-driven features. The input image is first mapped to bucket values via classical convolution, with kernels constrained to physically meaningful speckle patterns. These bucket values are then fed into hardware-efficient variational quantum circuits with patch encoding and star-topology entanglement for high-order feature extraction.

The physical measurement process of ghost imaging is expressed as
\begin{equation}
B_i=\sum_{x}\sum_{y} I_i(x,y)T(x,y),
\label{eq1}
\end{equation}
or in matrix form $\mathbf{B} = \mathbf{A} \mathbf{T}$, where $\mathbf{A} \in \mathbb{R}^{M \times N}$ is the speckle matrix, $\mathbf{T} \in \mathbb{R}^{N}$ denotes the target object vector, and $\mathbf{B} \in \mathbb{R}^{M}$ is the bucket vector.

Conventional recognition tasks are typically formulated as
\begin{equation}
\mathbf{T}^*=\arg\min_{\mathbf{T}}\frac12\|\mathbf{B}-\mathbf{A}\mathbf{T}\|_2^2+\lambda R_{\rm speckle}(\mathbf{A}),
\label{eq2}
\end{equation}
where the second term constrains the physical plausibility of speckles. However, fixing $\mathbf{A}$ and optimizing only $\mathbf{T}$ under low-dimensional measurements often overfits noise and fails to yield discriminative features at low sampling rates.

Leveraging superposition and entanglement, quantum circuits offer richer representation capability and global correlation modeling within a compact parameter space compared with classical networks. Moreover, the linear response of quantum feature mapping is highly compatible with the forward GI projection, enabling effective latent information extraction from low-dimensional bucket values.

Instead of explicitly optimizing $\mathbf{T}$, we introduce a variational quantum mapping $\Phi(\cdot;\Theta)$ to implicitly extract target features, and replace the fixed speckle matrix $\mathbf{A}$ with trainable convolution kernels $\mathbf{W}_{\rm conv}$, whose outputs are exactly the bucket values $\mathbf{B}(\mathbf{W}_{\rm conv})$. The revised objective is
\begin{equation}
\{W_{\rm conv}^*,\Theta^*\}
=\arg\min_{W_{\rm conv},\Theta}\frac12\big\|y_{\rm true}-f_{\rm cls}\big(\Phi\big(B(W_{\rm conv});\Theta\big)\big)\big\|_2^2
+\lambda\|W_{\rm conv}-S\|_F^2,
\label{eq3}
\end{equation}
where $\mathbf{S}$ denotes the initial reference speckle matrix and $f_{\rm cls}$ is a fully-connected classification mapping. This formulation breaks the two-stage design paradigm and allows adaptive speckle adjustment to maintain strong discrimination under low-sampling conditions.

\subsection{Quantum Neural Network}

\begin{figure}[!t]
\includegraphics[width=.7\textwidth]{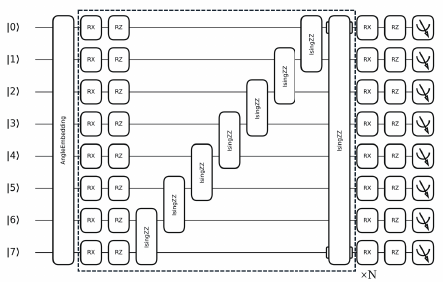}
\caption{\label{FIG2}Detailed architecture of the parameterized quantum circuit.}
\end{figure}

The quantum feature extraction layer exploits superposition and entanglement to capture high-order correlations from patch-preprocessed data. To comply with NISQ device constraints and enhance local feature extraction, we design a parameterized quantum circuit (ansatz), illustrated in Fig.~\ref{FIG2}

We first encode the input bucket vector into quantum states via a patch strategy. Given a vector of dimension $M$, the patch strategy partitions it into $n = M/m$ non-overlapping sub-blocks, each containing $m$ bucket values:
\begin{equation}
\mathbf{x}^k = (x_1^k, x_2^k, \dots, x_m^k), \quad \mathbf{x}^k \in \mathbb{R}^m,
\label{eq4}
\end{equation}
where $k = 1, 2, \dots, n$. Each sub-block $\mathbf{x}^k$ is encoded into $m$ qubits via angle embedding. After normalization to $[0,\pi]$, the data are loaded as
\begin{equation}
|\psi_k^{\text{in}}\rangle = \mathcal{E}_{\text{angle}}(\mathbf{x}^k),
\label{eq5}
\end{equation}
where $\mathcal{E}_{\text{angle}}(\cdot)$ denotes the angle embedding mapping that transforms a classical vector into an $m$-qubit state by encoding each entry as a rotation angle applied to the initial state $|0\rangle^{\otimes m}$.

The quantum feature extractor consists of alternating single-qubit rotation layers and two-qubit entanglement layers. The single-qubit rotation is defined as
\begin{equation}
R^{(k)}(\theta,\phi) = R_x(\theta) R_z(\phi),
\label{eq6}
\end{equation}
with $R_x(\theta)=e^{-i\theta\sigma_x/2}$ and $R_z(\phi)=e^{-i\phi\sigma_z/2}$. The entanglement layer adopts a star-topology Ising-$ZZ$ structure:
\begin{equation}
E^{(k)}(\gamma) = \prod_{\{p,q\}\in E_*} \text{IsingZZ}(\gamma_{pq}),
\label{eq7}
\end{equation}
where $\text{IsingZZ}(\gamma_{pq}) = \exp(-i\gamma_{pq} Z_p Z_q / 2)$, and $E_*$ denotes the set of edges connecting the central qubit to all other qubits in the block.

A basic module composed of one entanglement layer and one rotation layer is repeated $N$ times. The $r$-th layer operation of the $k$-th block is
\begin{equation}
L_r^{(k)} = E^{(k)}(\gamma_r) R^{(k)}(\theta_r, \phi_r).
\label{eq8}
\end{equation}

After $N$ repetitions, an additional single-qubit rotation is appended. The unitary of the $k$-th parallel circuit is
\begin{equation}
U_k(\Theta^{(k)}) = R^{(k)}(\theta_{\text{out}}, \phi_{\text{out}}) \prod_{r=1}^{N} L_r^{(k)}.
\label{eq9}
\end{equation}

The total ansatz is the tensor product of all independent subcircuits:
\begin{equation}
U(\Theta) = U_1 \otimes U_2 \otimes \cdots \otimes U_n,
\label{eq10}
\end{equation}
where $\Theta$ collects all trainable parameters.

After quantum evolution, each qubit is measured in the Pauli-$Z$ basis. The expectation value of the $i$-th qubit in block $k$ is
\begin{equation}
\langle Z_{k,i} \rangle = \langle \Psi_k^{\text{out}} | Z_i | \Psi_k^{\text{out}} \rangle, \quad
|\Psi_k^{\text{out}}\rangle = U_k(\Theta^{(k)}) |\psi_k^{\text{in}}\rangle.
\label{eq11}
\end{equation}

All expectation values are concatenated into a global quantum feature vector:
\begin{equation}
\mathbf{y}_{\text{out}} = \big( \langle Z_{1,1} \rangle, \dots, \langle Z_{1,m} \rangle, \langle Z_{2,1} \rangle, \dots, \langle Z_{n,m} \rangle \big)^{\mathrm{T}} \in \mathbb{R}^M.
\label{eq12}
\end{equation}

Through the above parallel block-wise quantum circuit for quantum state evolution and Pauli-$Z$ basis measurement, we can concatenate all measured expectations to obtain the global quantum feature vector. The full forward propagation process of the patch-based star-entangled quantum circuit, together with the quantum gradient calculation procedure using the parameter-shift rule, are summarized in Algorithm~\ref{alg:qcircuit}.

\begin{algorithm}[!t]
\caption{Patch-based Star-Entangled Quantum Circuit Forward \& Gradient Calculation}
\label{alg:qcircuit}
\SetKwInOut{Input}{Input}
\SetKwInOut{Output}{Output}
\Input{Bucket signal vector $\mathbf{B}\in\mathbb{R}^M$; Number of patches $n$, patch size $m$; Circuit depth $N$; Trainable quantum parameters $\Theta = \{\theta_r,\phi_r,\gamma_r,\theta_{\text{out}},\phi_{\text{out}}\}_{r=1}^N$.}
\Output{Quantum feature vector $\mathbf{y}_{\text{out}}$; Gradient $\partial \mathcal{L}/\partial \Theta$ for quantum parameter update.}

\textbf{Patch partition}: Split $\mathbf{B}$ into $n$ sub‑blocks $\{\mathbf{x}^1,\mathbf{x}^2,\dots,\mathbf{x}^n\}$;

\For{$k = 1$ \KwTo $n$}{
    Normalize $\mathbf{x}^k \leftarrow [0,\pi]$;
    Angle embedding: $|\psi_k^{\text{in}}\rangle \leftarrow \mathcal{E}_{\text{angle}}(\mathbf{x}^k)$;
    Initialize block unitary $U_k \leftarrow \mathbb{I}^{\otimes m}$;
    \For{$r = 1$ \KwTo $N$}{
        Single‑qubit rotation layer: $R^{(k)}(\theta_r,\phi_r) \leftarrow R_x(\theta_r)R_z(\phi_r)$;
        Star entanglement layer: $E^{(k)}(\gamma_r) \leftarrow \prod\limits_{\{p,q\}\in E_*}\operatorname{IsingZZ}(\gamma_{pq})$;
        Combine layer: $L_r^{(k)} \leftarrow E^{(k)}(\gamma_r) R^{(k)}(\theta_r, \phi_r)$;
        Update block unitary: $U_k \leftarrow L_r^{(k)} U_k$;
    }
    Append output rotation: $U_k \leftarrow R^{(k)}(\theta_{\text{out}},\phi_{\text{out}}) U_k$;
    Quantum evolution: $|\Psi_k^{\text{out}}\rangle \leftarrow U_k |\psi_k^{\text{in}}\rangle$;
    Measure Pauli‑$Z$ expectation values $\langle Z_{k,1}\rangle,\dots,\langle Z_{k,m}\rangle$;
}

Concatenate all block expectations to form global feature:
$\mathbf{y}_{\text{out}} \leftarrow \big(\langle Z_{1,1}\rangle,\dots,\langle Z_{n,m}\rangle\big)^{\mathrm{T}}$;

Compute quantum gradients via parameter‑shift rule:
$\displaystyle\frac{\partial \mathcal{L}}{\partial \theta} = \frac{\mathcal{L}(\theta+\pi/2)-\mathcal{L}(\theta-\pi/2)}{2},\quad \forall\theta\in\Theta$;

\Return{$\mathbf{y}_{\text{out}},\ \partial\mathcal{L}/\partial\Theta$}
\end{algorithm}

The final classification output is obtained via a fully-connected layer with Softmax activation:
\begin{equation}
\mathbf{y}_{\text{pred}} = \text{Softmax}(\mathbf{W} \mathbf{y}_{\text{out}} + \mathbf{b}), \quad
\mathbf{W} \in \mathbb{R}^{C \times M}, \; \mathbf{b} \in \mathbb{R}^C,
\label{eq13}
\end{equation}
where $\mathbf{W}$ and $\mathbf{b}$ are the weight matrix and bias vector, and $C$ is the number of classes.

The discrete loss function corresponding to Eq.~(\ref{eq3}) is
\begin{equation}
\mathcal{L} = \sum_{b=1}^{B} \sum_{c=1}^{C} \big( y_{\text{true}}^{(b,c)} - y_{\text{pred}}^{(b,c)} \big)^2
+ \lambda \sum_{k=1}^{K} \| \mathbf{W}_{\text{conv}}^{(k)} - \mathbf{S}^{(k)} \|_F^2,
\label{eq14}
\end{equation}
where the first term is the supervised loss and the second term enforces speckle consistency regularization. Since the model contains both classical and quantum parameters, separate update rules are applied:
\begin{equation}
\mathbf{W}_{\text{conv}} \leftarrow \mathbf{W}_{\text{conv}} - \alpha \frac{\partial \mathcal{L}}{\partial \mathbf{W}_{\text{conv}}}, \qquad
\Theta \leftarrow \Theta - \alpha \frac{\partial \mathcal{L}}{\partial \Theta}.
\label{eq15}
\end{equation}
where $\mathbf{W}_{\text{conv}}$ is calculated through backpropagation, and $\Theta$ is obtained using parameter translation rules.

This hybrid optimization strategy can alleviate the instability of quantum feature extraction at low sampling rates while ensuring physical consistency, ultimately improving the robustness of the overall recognition. Algorithm~\ref{alg:sq_hybrid} summarizes the complete process of the proposed recognition method.

\begin{algorithm}[!t]
\caption{Hybrid Recognition}
\label{alg:sq_hybrid}
\SetKwInOut{Input}{Input}
\SetKwInOut{Output}{Output}
\Input{Training dataset $\{X^{(b)}, \mathbf{y}_{\text{true}}^{(b)}\}_{b=1}^{B}$; Initial reference speckle matrix $\mathbf{S}$; Regularization coefficient $\lambda$; Learning rate $\alpha$; Maximum iterations $T$; Loss tolerance $\tau$.}
\Output{Optimized convolution kernels $\mathbf{W}_{\text{conv}}^{*}$; Quantum circuit parameters $\Theta^{*}$; Classifier parameters $\{\mathbf{W}, \mathbf{b}\}$.}

\textbf{Initialize}:
$\mathbf{W}_{\text{conv}} \leftarrow \mathbf{S}$;
$\Theta \leftarrow \text{Random}$;
$\{\mathbf{W}, \mathbf{b}\} \leftarrow \text{Random}$;
$t \leftarrow 0$;
$\mathcal{L}^{(0)} \leftarrow +\infty$;

\While{$\mathcal{L}^{(t)} > \tau$ and $t \leq T$}{
    Compute bucket signals: $\mathbf{B}^{(b)} \leftarrow \mathbf{B}(\mathbf{W}_{\text{conv}})$;
    Infer predictions: $\mathbf{y}_{\text{pred}}^{(b)} \leftarrow f_{\text{cls}}\bigl(\Phi(\mathbf{B}^{(b)};\Theta)\bigr)$;
    Compute loss:
    \[
    \mathcal{L}^{(t)} \leftarrow \sum_{b,c}\bigl(y_{\text{true}}^{(b,c)}-y_{\text{pred}}^{(b,c)}\bigr)^{2}
    + \lambda\|\mathbf{W}_{\text{conv}}-\mathbf{S}\|_{F}^{2};
    \]
    Update parameters:
    \[
    \mathbf{W}_{\text{conv}} \leftarrow \mathbf{W}_{\text{conv}} - \alpha \frac{\partial\mathcal{L}}{\partial\mathbf{W}_{\text{conv}}},\quad
    \Theta \leftarrow \Theta - \alpha \frac{\partial\mathcal{L}}{\partial\Theta};
    \]
    $t \leftarrow t + 1$;
}
\Return{$\mathbf{W}_{\text{conv}}^{*}, \Theta^{*}, \{\mathbf{W}, \mathbf{b}\}$}
\end{algorithm}

\section{Results and Discussion}
\label{sec:results}

We evaluated the proposed method through numerical simulations and optical experiments. All simulations were performed on a workstation equipped with a 13th-generation Intel Core i9-13900H processor and an NVIDIA GeForce RTX 4070 GPU. The MNIST dataset (28$\times$28) was scaled down to 32$\times$32, with 25,000 samples used for training and 10,000 samples used for testing. The optimizer was Adam, with a batch size of 64, a learning rate of 0.001, and a regularization coefficient $\lambda=1.0$, for a total of 100 epochs. Quantum circuit simulations were implemented using the PennyLane and TensorCircuit frameworks in Python.

\begin{figure}[!t]
\includegraphics{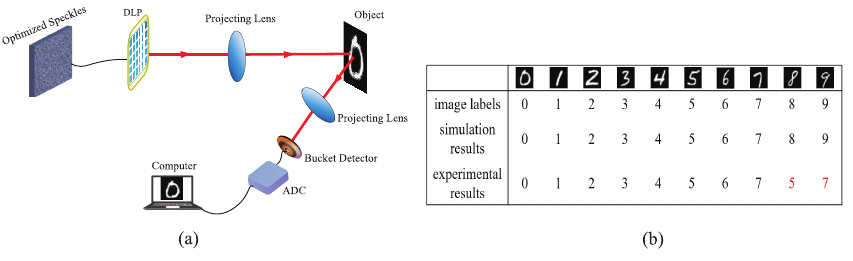}
\caption{\label{FIG3}Recognition results of different digits. (a) Experimental setup; (b) Simulated and experimental output.}
\end{figure}

Fig.~\ref{FIG3}(a) shows a schematic diagram of the optical setup used for collecting bucket values in the experiment. First, the optimized convolutional kernel is loaded onto a TI DLP LightCrafter 4500 digital light projector, which generates a speckle pattern at a 120 Hz refresh rate and 8-bit grayscale. The speckle pattern is then projected onto the target object, and the reflected light is collected by a bucket detector (PMM02-1, ThorLabs). The collected analog signal is digitized via a USB-6341 data acquisition card (National Instruments). Finally, the collected bucket values are input into a trained quantum network for identification. It should be noted that once the actual bucket measurements are obtained from the detector, all subsequent classification calculations are performed on a classical computer based on the trained quantum model.

Fig.~\ref{FIG3}(b) compares the results of ten categories of handwritten digit recognition obtained from simulation (based on synthesized bucket values) and experiments (based on real optical measurements). Under simulation conditions, the proposed method achieves perfect classification of all digits; however, under experimental conditions, the proposed method misclassifies the digits 8 and 9. The uneven surface reflectivity, stray light interference, and photoelectric conversion loss that may occur under the hardware environment weaken the ability to distinguish between the double-ring structure of the digit `8' and the bottom curvature features of `9', thus leading to a decrease in recognition performance in the experiment. Under non-ideal conditions, given the low information content of ghost imaging measurements themselves, the quantum feature extraction layer's ability to distinguish structurally similar digits will decrease.

\begin{figure}[!t]
\includegraphics{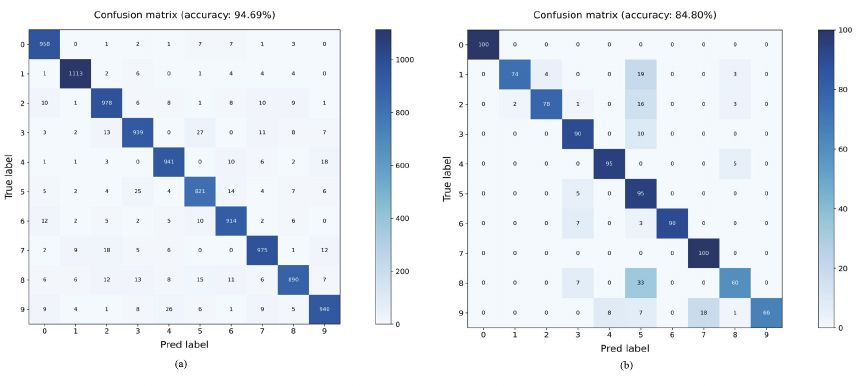}
\caption{\label{FIG4}Confusion matrices for the test set. (a) Simulation; (b) Experiment.}
\end{figure}

Fig.~\ref{FIG4} shows the confusion matrices from simulation and experiments. The simulation matrix reveals that diagonal elements dominate, confirming the superior inter-class discrimination ability of the proposed method under ideal conditions. In contrast, the experimental matrix exhibits significant off-diagonal confusion: the digit 8 is frequently misclassified as 5, and the digit 9 is frequently misclassified as 7---a pattern consistent with the feature degradation analysis caused by the aforementioned hardware environment. This comparison clearly demonstrates the impact of real-world noise on recognition performance and highlights the necessity of joint speckle-quantum optimization for robust practical deployment.

\begin{figure}[!t]
\includegraphics[width=.7\textwidth]{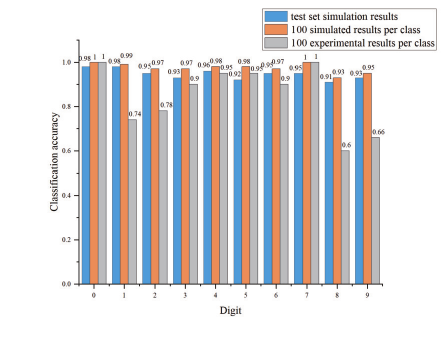}
\caption{\label{FIG5}Classification accuracy per digit in the test set.}
\end{figure}

Fig.~\ref{FIG5} shows the number-by-number classification accuracy for three evaluation sets: the full simulation test set (blue), the simulation subset with 100 samples per class (orange), and the experimental subset with 100 samples per class (grey). In the full simulation test set, the accuracy for each class ranged from 0.93 (number 3) to 0.99 (number 0), with an average accuracy of 0.945. The simulation subset with 100 samples per class performed slightly better, with an accuracy ranging from 0.94 to 0.99 and an average of 0.965. In contrast, the experimental subset showed a significant drop in performance, with an average accuracy of only 0.84. Digits 8 and 9 are the most severely affected (0.60 and 0.66, respectively), followed by digits 2 and 3 (0.74 and 0.78), while digits 0, 1, 4, 5, 6, and 7 maintain accuracies above 0.90. The results above demonstrate that the proposed method remains robust to most digit categories, even in real-world noisy environments.

We compared our proposed method with two baseline models: classical convolutional neural networks (CML) \cite{lyu2017deep} and traditional hybrid quantum machine learning models (QML) \cite{xiao2023practical}. The number of parameters in both baselines is comparable to that in our proposed model.

\begin{figure}[!t]
\includegraphics[width=.7\textwidth]{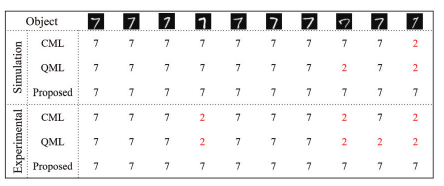}
\caption{\label{FIG6}Classification results for digit ``7''.}
\end{figure}

Fig.~\ref{FIG6} visualizes the classification results of the three models on the digit 7. It is evident that while the classic CML model incorporates physical priors, its final discrimination performance is poor due to limitations in feature extraction capabilities. Traditional QML, employing fixed speckle patterns, cannot selectively enhance key structural features of the target, resulting in missed detections and the worst performance. In contrast, our joint optimization model enables the speckle pattern and quantum circuitry to co-evolve, adaptively enhancing the discriminative features of the digit 7 (such as the upper right corner bend and vertical stroke structure), thus achieving accurate classification. The above comparison demonstrates that, under ghost imaging conditions, our architecture possesses a unique advantage in preserving weak discriminative features.

\begin{table}[!t]
\footnotesize
\caption{\label{tab:comparison}Performance comparison of three models on MNIST and Fashion-MNIST (F-MNIST) datasets under different sampling rates.}
\begin{ruledtabular}
\begin{tabular}{llddd}
Dataset & Method &
\multicolumn{1}{c}{1.5625\%} &
\multicolumn{1}{c}{3.125\%} &
\multicolumn{1}{c}{6.25\%} \\
\colrule
MNIST   & CML      & 0.878 & 0.926 & 0.942 \\
        & QML      & 0.789 & 0.904 & 0.930 \\
        & Proposed & 0.901 & 0.947 & 0.979 \\
\colrule
F-MNIST & CML      & 0.780 & 0.860 & 0.880 \\
        & QML      & 0.711 & 0.800 & 0.840 \\
        & Proposed & 0.817 & 0.873 & 0.904 \\
\end{tabular}
\end{ruledtabular}
\end{table}

Table~\ref{tab:comparison} quantitatively compares the performance of the three methods on the MNIST and Fashion-MNIST datasets at sampling rates of $1.5625\%$, $3.125\%$, and $6.25\%$. The proposed method consistently outperforms both baselines across all settings, validating the effectiveness of the joint speckle-quantum optimization. On the MNIST dataset, the proposed method achieves an accuracy of 0.901 at the lowest sampling rate ($1.5625\%$), outperforming CML and QML by 0.023 and 0.112, respectively; the accuracy further increases to 0.979 when the sampling rate is increased to $6.25\%$. On the more challenging Fashion-MNIST dataset, the proposed method achieves accuracies of 0.817, 0.873, and 0.904 at the three sampling rates mentioned above. These results demonstrate that the proposed method architecture effectively alleviates the information scarcity problem under low sampling conditions and overcomes the performance bottleneck of traditional two-stage methods.

\section{Conclusion}
\label{sec:conclusion}

We have proposed a ghost imaging recognition method based on joint optimization of speckle patterns and quantum network parameters. By introducing a speckle consistency regularization mechanism, end-to-end synchronous optimization of the speckle structure and the variable quantum feature space is achieved. Simultaneously, a block-parallel strategy is employed to adapt to the hardware constraints of noisy medium-scale quantum (NISQ) devices.

Simulation results show that, at an ultra-low sampling rate of $1.5625\%$, the proposed method achieves classification accuracies of $90.1\%$ and $81.7\%$ on the MNIST and Fashion-MNIST datasets, respectively, significantly outperforming the baseline model under the same experimental conditions, and exhibiting strong robustness to quantum noise. Physical experiments further validate the effectiveness of the framework, with an overall average recognition accuracy of $84.8\%$, and an accuracy exceeding $90\%$ for digits with relatively simple topologies. These results confirm that the proposed physics--quantum fusion architecture effectively addresses information scarcity and enhances recognition performance under low-sampling and hardware-constrained scenarios.

\begin{acknowledgments}
This work was supported by the National Natural Science Foundation of China (Grant No. 62375140, 61871234), the Postgraduate Research \& Practice Innovation Program of Jiangsu Province (KYCX24-1191).
\end{acknowledgments}

\bibliography{reference}

@article{strekalov1995observation,
  title={Observation of two-photon “ghost” interference and diffraction},
  author={Strekalov, DV and Sergienko, AV and Klyshko, DN and Shih, YH},
  journal={Physical review letters},
  volume={74},
  number={18},
  pages={3600},
  year={1995},
  publisher={APS}
}

@article{pittman1995optical,
  title={Optical imaging by means of two-photon quantum entanglement},
  author={Pittman, Todd B and Shih, Yanhua H and Strekalov, Dmitry V and Sergienko, Alexander V},
  journal={Physical Review A},
  volume={52},
  number={5},
  pages={R3429},
  year={1995},
  publisher={APS}
}

@article{bennink2002two,
  title={“Two-photon” coincidence imaging with a classical source},
  author={Bennink, Ryan S and Bentley, Sean J and Boyd, Robert W},
  journal={Physical review letters},
  volume={89},
  number={11},
  pages={113601},
  year={2002},
  publisher={APS}
}

@article{shapiro2008computational,
  title={Computational ghost imaging},
  author={Shapiro, Jeffrey H},
  journal={Physical Review A—Atomic, Molecular, and Optical Physics},
  volume={78},
  number={6},
  pages={061802},
  year={2008},
  publisher={APS}
}

@article{howland2011photon,
  title={Photon-counting compressive sensing laser radar for 3D imaging},
  author={Howland, Gregory A and Dixon, P Ben and Howell, John C},
  journal={Applied optics},
  volume={50},
  number={31},
  pages={5917--5920},
  year={2011},
  publisher={Optical Society of America}
}

@article{yang2023underwater,
  title={Underwater environment laser ghost imaging based on Walsh speckle patterns},
  author={Yang, Mochou and Wu, Yi and Feng, Guoying},
  journal={Frontiers in Physics},
  volume={11},
  pages={1106320},
  year={2023},
  publisher={Frontiers Media SA}
}

@article{yang2024ghost,
  title={A ghost imaging framework based on laser mode speckle pattern for underwater environments},
  author={Yang, Mo-Chou and Wang, Peng and Wu, Yi and Feng, Guo-Ying},
  journal={Communications Engineering},
  volume={3},
  number={1},
  pages={52},
  year={2024},
  publisher={Nature Publishing Group UK London}
}

@article{darafsheh2012optical,
  title={Optical super-resolution by high-index liquid-immersed microspheres},
  author={Darafsheh, Arash and Walsh, Gary F and Dal Negro, Luca and Astratov, Vasily N},
  journal={Applied Physics Letters},
  volume={101},
  number={14},
  year={2012},
  publisher={AIP Publishing}
}

@article{ordonez2024single,
  title={Single-pixel microscopy with optical sectioning},
  author={Ord{\'o}{\~n}ez, Luis and Lenz, Armin JM and Ipus, Erick and Lancis, Jes{\'u}s and Tajahuerce, Enrique},
  journal={Optics Express},
  volume={32},
  number={15},
  pages={26038--26051},
  year={2024},
  publisher={Optica Publishing Group}
}

@article{NJYD202103009,
author = {Cui, Qiang and Min, Lihua and Shi, Youxin},
title = {Image Segmentation Method Based on Retinex Theory and Local Gray Information},
journal = {Journal of Nanjing University of Posts and Telecommunications (Natural Science Edition)},
volume = {41},
number = {3},
pages = {62--71},
year = {2021},
issn = {1673-5439},
doi = {10.14132/j.cnki.1673-5439.2021.03.009}
}

@article{chen2025attention,
  title={Attention-enhanced computational ghost imaging},
  author={Chen, Yifan and Tian, Tong and Lu, Xin and Li, Chen and Zhu, Ruolan and Sun, Zhe and Li, Xuelong},
  journal={Science China Information Sciences},
  volume={68},
  number={6},
  pages={162104},
  year={2025},
  publisher={Springer}
}

@article{zhang2026optical,
  title={Optical-chaos temporal ghost imaging},
  author={Zhang, Rong and Wang, Anbang and Wang, Longsheng and Jia, Zhiwei and Wang, Yuncai and Qin, Yuwen},
  journal={Science China Information Sciences},
  volume={69},
  number={6},
  pages={162407},
  year={2026},
  publisher={Springer}
}

@article{lyu2017deep,
  title={Deep-learning-based ghost imaging},
  author={Lyu, Meng and Wang, Wei and Wang, Hao and Wang, Haichao and Li, Guowei and Chen, Ni and Situ, Guohai},
  journal={Scientific reports},
  volume={7},
  number={1},
  pages={17865},
  year={2017},
  publisher={Nature Publishing Group UK London}
}

@article{he2025ghost,
  title={Ghost edge detection based on joint optimization of speckle patterns and network architecture},
  author={He, Yunfan and Ma, Yuanyuan and Wang, Le and Zhao, Shengmei},
  journal={Optics \& Laser Technology},
  volume={192},
  pages={114047},
  year={2025},
  publisher={Elsevier}
}

@article{pan2022quantum,
  title={Quantum algorithm for neighborhood preserving embedding},
  author={Pan, Shi-Jie and Wan, Lin-Chun and Liu, Hai-Ling and Wu, Yu-Sen and Qin, Su-Juan and Wen, Qiao-Yan and Gao, Fei},
  journal={Chinese Physics B},
  volume={31},
  number={6},
  pages={060304},
  year={2022},
  publisher={Chinese Physical Society and IOP Publishing Ltd}
}

@article{lamichhane2025quantum,
  title={Quantum machine learning: Recent advances, challenges, and perspectives},
  author={Lamichhane, Pradeep and Rawat, Danda B},
  journal={IEEE Access},
  volume={13},
  pages={94057--94105},
  year={2025},
  publisher={IEEE}
}

@article{xu2022scalable,
  title={Scalable multiple GHZ states equations and its applications in efficient quantum key agreement},
  author={Xu, Yuguang and Wang, Chaonan and Wang, Xueying and Zhu, Hongfeng},
  journal={Quantum Information Processing},
  volume={21},
  number={3},
  pages={91},
  year={2022},
  publisher={Springer}
}

@article{tacchino2020quantum,
  title={Quantum implementation of an artificial feed-forward neural network},
  author={Tacchino, Francesco and Barkoutsos, Panagiotis and Macchiavello, Chiara and Tavernelli, Ivano and Gerace, Dario and Bajoni, Daniele},
  journal={Quantum Science \& Technology},
  volume={5},
  number={4},
  pages={044010},
  year={2020},
  publisher={IOP Publishing}
}

@article{henderson2020quanvolutional,
  title={Quanvolutional neural networks: powering image recognition with quantum circuits},
  author={Henderson, Maxwell and Shakya, Samriddhi and Pradhan, Shashindra and Cook, Tristan},
  journal={Quantum Machine Intelligence},
  volume={2},
  number={1},
  pages={2},
  year={2020},
  publisher={Springer}
}

@article{zhang2023single,
  title={Single-qubit quantum classifier based on gradient-free optimization algorithm},
  author={Zhang, Anqi and Wang, Kelun and Wu, Yihua and Zhao, Sheng-Mei},
  journal={Chinese Physics B},
  volume={32},
  number={10},
  pages={100308},
  year={2023},
  publisher={Chinese Physical Society and IOP Publishing Ltd}
}

@article{havlivcek2019supervised,
  title={Supervised learning with quantum-enhanced feature spaces},
  author={Havl{\'\i}{\v{c}}ek, Vojt{\v{e}}ch and C{\'o}rcoles, Antonio D and Temme, Kristan and Harrow, Aram W and Kandala, Abhinav and Chow, Jerry M and Gambetta, Jay M},
  journal={Nature},
  volume={567},
  number={7747},
  pages={209--212},
  year={2019},
  publisher={Nature Publishing Group UK London}
}

@article{hur2022quantum,
  title={Quantum convolutional neural network for classical data classification},
  author={Hur, Tak and Kim, Leeseok and Park, Daniel K},
  journal={Quantum Machine Intelligence},
  volume={4},
  number={1},
  pages={3},
  year={2022},
  publisher={Springer}
}

@article{li2022recent,
  title={Recent advances for quantum classifiers},
  author={Li, Weikang and Deng, Dong-Ling},
  journal={Science China Physics, Mechanics \& Astronomy},
  volume={65},
  number={2},
  pages={220301},
  year={2022},
  publisher={Springer}
}

@article{xiao2023practical,
  title={Practical advantage of quantum machine learning in ghost imaging},
  author={Xiao, Tailong and Zhai, Xinliang and Wu, Xiaoyan and Fan, Jianping and Zeng, Guihua},
  journal={Communications Physics},
  volume={6},
  number={1},
  pages={171},
  year={2023},
  publisher={Nature Publishing Group UK London}
}

@article{zhai2025quantum,
  title={Quantum neural compressive sensing for ghost imaging},
  author={Zhai, Xinliang and Xiao, Tailong and Huang, Jingzheng and Fan, Jianping and Zeng, Guihua},
  journal={Physical Review Applied},
  volume={23},
  number={1},
  pages={014018},
  year={2025},
  publisher={APS}
}

@article{li2024quantum,
  title={Quantum self-attention neural networks for text classification},
  author={Li, Guangxi and Zhao, Xuanqiang and Wang, Xin},
  journal={Science China Information Sciences},
  volume={67},
  number={4},
  pages={142501},
  year={2024},
  publisher={Springer}
}

@article{zhang55special,
  title={Special Topic: Quantum Information},
  author={ZHANG, Weiping and ZHOU, Zhengwei and SU, Xiaolong and XU, Jinshi},
  journal={SCIENTIA SINICA Informationis},
  volume={55},
  number={9},
  pages={2400},
  year={2025},
  publisher={Science China Press}
}

\end{document}